\documentclass[apj]{emulateapj}
\usepackage{amsfonts}

\newcommand{\etal}{{et al.~}}
\newcommand{\msunh}{\>h^{-1}\rm M_\odot}
\newcommand{\Msun}{\>{\rm M_\odot}}

\newcommand{\lta}{\la}
\newcommand{\mpch}{\>h^{-1}{\rm {Mpc}}}

\newcommand{\kpc}{\>{\rm kpc}}

\shorttitle{Alignment of groups}

\shortauthors{Wang et al.}

\begin{document}

\title{Alignments of Group Galaxies with Neighboring Groups}
\author{Yougang Wang\altaffilmark{1}, Changbom Park\altaffilmark{2}, Xiaohu
  Yang\altaffilmark{3}, Yun-Young Choi\altaffilmark{4}, Xuelei
  Chen\altaffilmark{1}}

\altaffiltext{1}{National Astronomical Observatories, Chinese Academy of
  Sciences, Beijing 100012, China; E-mail: wangyg@bao.ac.cn}
\altaffiltext{2}{Korea Institute for Advanced Study, Dongdaemun-gu, Seoul
  130-722, Korea; cbp@kias.re.kr}
\altaffiltext{3}{Key Laboratory for Research in Galaxies and Cosmology,
  Shanghai Astronomical Observatory, the Partner Group of MPA, Nandan Road 80,
  Shanghai 200030, China}
\altaffiltext{4}{Astrophysical Research Center for the Structure and Evolution
  of the Cosmos, Sejong University, Seoul 143-747, Korea}


\begin{abstract} Using  a sample of galaxy  groups found in  the Sloan Digital
  Sky Survey Data Release 4, we  measure the following four types of alignment
  signals: (1)  the alignment between  the distributions of the  satellites of
  each group  relative to the direction  of the nearest  neighbor group (NNG);
  (2) the alignment between the major  axis direction of the central galaxy of
  the host group (HG) and the  direction of the NNG; (3) the alignment between
  the major axes  of the central galaxies of  the HG and the NNG;  and (4) the
  alignment  between the  major  axes of  the  satellites of  the  HG and  the
  direction  of the  NNG.   We find  strong  signal of  alignment between  the
  satellite distribution and the orientation of central galaxy relative to the
  direction of the NNG, even when  the NNG is located beyond $3r_{\rm vir}$ of
  the host group.  The  major axis of the central galaxy of  the HG is aligned
  with the direction of the NNG.  The alignment signals are more prominent for
  groups that are  more massive and with early type  central galaxies. We also
  find  that there  is a  preference for  the two  major axis  of  the central
  galaxies of  the HG and NNG  to be parallel  for the system with  both early
  central galaxies,  however not for the  systems with both  late type central
  galaxies.  For  the orientation  of satellite galaxies,  we do not  find any
  significant alignment  signals relative to  the direction of the  NNG.  From
  these four types of alignment measurements, we conclude that the large scale
  environment traced  by the nearby group  affects primarily the  shape of the
  host dark matter halo, and  hence also affects the distribution of satellite
  galaxies  and the  orientation of  central galaxies.   In addition,  the NNG
  directly  affects the  distribution of  the satellite  galaxies  by inducing
  asymmetric  alignment signals.  And NNG  at very  small separation  may also
  contribute a second  order impacts on the orientation  of the central galaxy
  in the HG.
\end{abstract}


\section{Introduction}

The distribution of satellites in the groups of galaxies holds
important clues to the  assembly history  of dark matter  halos.
Since satellite  galaxies are typically  distributed over the entire
dark  matter halo,  they are  a useful tracer  of  the  dark matter
distribution  on  the  scale  of the  group.  In particular, their
position  provides information  on  the shape  of the  dark matter
halo (Carter \& Metcalfe 1980; Plionis, Barrow \& Frenk 1991; Fasano
et al. 1993; Basilakos, Plionis \& Maddox 2000; Orlov, Petrova \&
Martynova 2001; Plionis  et al.  2004,  2006; Bailin  \& Steinmetz
2005;  Wang  et al.  2008, hereafter W08), and their kinematics
could be used to estimate the mass of the haloes  (e.g., Zaritsky et
al.  1993, 1997;  McKay et  al. 2002;  Brainerd \& Specian  2003;
Katgert, Biviano \& Mazure 2004;  van den  Bosch  et al.  2004; More
et al. 2009a, 2009b).

One way to characterize  quantitatively the distribution of satellite galaxies
is  to  measure the  alignment  between  their  spatial distribution  and  the
orientation of their central galaxies.  Extensive studies with high-resolution
simulations have  shown that sub-haloes tend  to align with the  major axis of
their host  halos (Knebe et  al. 2004, 2008a,  2008b; Libeskind et  al.  2005,
2007;  Wang  et al.  2005;  Zentner  et al.  2005;  Kang  et  al.  2007).  The
observational search for  a possible alignment between the  central galaxy and
satellites  has a  long and  serpentine history.  The first  study of  such an
alignment  was performed  by Holmberg  (1969), who  found that  satellites are
preferentially   located    along   the   minor   axes    of   isolated   disc
galaxies.  Holmberg's  study  was  restricted to  projected  satellite-central
distances of $r_p  \lta 50 \kpc$. Subsequent studies,  however, were unable to
confirm this so-called ``Holmberg effect'' (Hawley \& Peebles 1975; Sharp, Lin
\& White 1979;  MacGillivray et al. 1982). Zaritsky et  al. (1997) studied the
distribution of satellites around spiral  hosts and were also unable to detect
any significant alignment for $r_p \lta  200 \kpc$, but they found a preferred
minor-axis alignment for $300 \kpc \lta  r_p \lta 500 \kpc$. The satellites of
our Milk Way galaxy  and the nearby M31 galaxy lie in  planes which are highly
inclined with respect  to their discs.  This were  noted by Lynden-Bell (1976,
1982),  Majewski (1994),  Hartwick (1996,  2000)  and Kroupa,  Theis \&  Boily
(2005) for  the Milky Way, by Koch  \& Grebel (2006) and  McConnachie \& Irwin
(2006) for M31, and by Metz, Kroupa \& Jerjen (2007) for both galaxies.

With large redshift  surveys, such as the 2dF  Galaxy Redshift Survey (2dFGRS;
Colless  et  al.  2001) and  the  Sloan  Digital  Sky  Survey (SDSS;  York  et
al. 2000),  much larger samples  of galaxy groups  can be used  to investigate
this alignment problem.  Sales \& Lambas (2004; 2009) used a  set of 1498 host
galaxies  with  3079 satellites  from  the  2dFGRS,  and found  a  large-scale
alignment of the  satellites along the host major axes for  $300 \kpc \lta r_p
\lta 500 \kpc$.   Brainerd (2005) studied a sample  of isolated SDSS galaxies,
and found that the distribution of satellite galaxies is strongly aligned with
the major  axis of the disc host  galaxy. Yang et al.   (2006, hereafter Y06),
using a galaxy group catalogue similar to  the one used here, but based on the
SDSS Data Release  2 (DR2), studied the alignment signal as  a function of the
color of  the central and satellite  galaxies.  They found  that the alignment
strength  is strongest  between red  centrals  and red  satellites, while  the
satellite distribution  in systems  with a blue  central galaxy  is consistent
with being isotropic.  Y06 also found that the  alignment strength is stronger
in  more  massive haloes  and  at smaller  projected  radii  from the  central
galaxy. These results have  subsequently been confirmed by several independent
studies  (Donoso, O'Mill  \& Lambas  2006; Azzaro  et al.  2007;  Agustsson \&
Brainerd 2006,  2007; W08;  Steffen \& Valenzuela  2008; Bailin et  al. 2008).
These studies have focused on whether the satellites are distributed along the
major axis or minor axis of the central galaxy.

Besides the alignment  between the distribution of satellite  galaxies and the
orientation of their  central galaxy, other forms of  alignment have also been
studied. These  include the  alignment between neighboring  clusters (Binggeli
1982; West 1989; Plionis 1994),  between brightest cluster galaxies (BCGs) and
their parent clusters (Carter \&  Metcalfe 1980; Binggeli 1982; Struble 1990),
between  the orientation  of satellite  galaxies  and the  orientation of  the
cluster  (Dekel 1985; Plionis  et al.  2003), and  between the  orientation of
satellite galaxies  and the orientation of  the BCG (Struble  1990). Using the
same group catalogue as the one used here, Faltenbacher et al. (2007) examined
three  different  types of  intrinsic  galaxy  alignment  within groups:  halo
alignment between  the orientation of  the brightest group galaxies  (BGG) and
the  distribution of  its  satellite galaxies,  radial  alignment between  the
orientation of a satellite galaxy and the direction toward its BGG, and direct
alignment between the orientation of the  BGG and that of its satellites. They
found  that the  orientations  of red  satellites  are preferentially  aligned
radially  in the direction  of the  BGG. In  addition, they  found a  weak but
significant  indication  that  the  orientations  of  satellite  galaxies  are
directly  aligned   with  that  of   their  BGG  on   scales  $r<0.1R_{\mathrm
vir}$. Based on a cosmological $N$-body simulation, Faltenbacher et al. (2008)
analyzed  the spatial  and kinematic  alignments of  satellite halos  within 5
times the virial  radius of group-sized host halos. They  found that the tidal
forces on the large scales can gives  rise to a halo alignment out to at least
$5R_\mathrm{vir}$. This means that  the alignment signal is strongly dependent
on the large-scale environment. It is also found that the orientations of dark
matter halos can be related to their surrounding structures, such as filaments
and large-scale  walls (e.g., Faltenbacher et  al. 2002; Einasto  et al. 2003;
Avila-Reese et al.  2005; Hopkins, Bahcall \& Bode 2005;  Kasun \& Evard 2005;
Basilakos et al. 2006; Altay et al. 2006; Aragon-Calvo et al. 2006; Maulbetsch
et al. 2007; Ragone-Figueroa \& Plionis 2007; Hahn et al.  2007a, 2007b). Paz,
Stasyszyn \& Padilla (2008) used  numerical simulations and the real data from
the SDSS  Data Release  6 (DR6)  to study the  alignments between  the angular
momentum of individual objects and  the large-scale structure. They found that
the angular momentum of dark  matter haloes are preferentially oriented in the
direction perpendicular to the distribution of matter, and more massive haloes
show a higher degree of alignment. Okumura, Jing \& Li (2009) investigated the
correlation  between  the  orientation  of  giant  elliptical  galaxies,  they
measured the intrinsic ellipticity correlation function of 83773 SDSS luminous
red galaxies (LRGs) and found that there is a positive alignment between pairs
of the LRGs up to  $~30h^{-1}$Mpc scales. Recently, Faltenbacher et al. (2009)
used the SDSS  DR6 and the Millennium simulation  to determinate the alignment
between  galaxies   and  large-scale  structure,  and  found   that  there  is
significant  alignment  between  the  major  axes of  red  galaxies  with  the
surrounding large-scale structure.


In this paper, we aim to study the impacts of the nearest neighbor
groups (hereafter NNGs) and  possibly the large scale environments
beyond the NNGs (e.g., NNG has a higher probability to be
distributed along the direction of the  filament)  on the  various
properties  of  the central  and  satellite galaxies, i.e., on the
distribution  of satellite galaxies and the shapes of the central
and satellite galaxies.   Throughout this paper,  unless stated
otherwise, when  we refer to the  impacts of the NNG,  possible
impacts from the large  scale environments  are not excluded. For
our purposes, the following  alignment  signals are  measured:  (1)
the alignment between  the distributions of the satellites relative
to the direction of the NNG; (2) the alignment between the major
axis of  the central galaxy of the host group (HG) and the direction
of the NNG; (3)  the alignment between the two major axes of the
central galaxies of the HG and  the NNG; and (4) the alignment
between the major axes of the satellites of the  HG and the
direction of the NNG. Here the alignment signals (2), (3) and (4)
are measured to probe the impact of the NNG on the shapes of the
central and satellite galaxies, respectively.

This  paper  is  organized  as   follows.   In  Section  2,  we  describe  the
observational  data used for  this study.   Section 3  presents the  method to
quantify the alignment signal. Section 4 shows the results of various kinds of
alignment signals we  measured, and their dependence on  the morphology of the
central galaxies of  the HG and the NNG.  Section 5  summarize our results and
discuss various related issues.

\section{Observational Data Set}
\subsection{Groups: central and satellite galaxies}

The analysis  presented in this  paper is based  on the SDSS DR4  galaxy group
catalogue of Yang \etal (2007)\footnote{In  this paper, we refer to systems of
  galaxies as groups regardless of their richness, including isolated galaxies
  (i.e., systems with a single  member) and rich clusters of galaxies.}.  This
group catalogue is constructed by applying the halo-based group finder of Yang
\etal  (2005)  to  the   New  York  University  Value-Added  Galaxy  Catalogue
(NYU-VAGC;  see  Blanton  \etal  2005),   which  is  based  on  the  SDSS  DR4
(Adelman-McCarthy  \etal 2006). From  this catalogue  Yang \etal  selected all
galaxies in  the Main Galaxy Sample with  redshifts in the range  $0.01 \leq z
\leq 0.20$ and  with a redshift completeness greater than  0.7. This sample of
galaxies is  further divided  into three group  samples: sample I,  which only
uses the $362,356$ galaxies with  measured redshifts from the SDSS; sample II,
which also  includes $7091$ galaxies  with SDSS photometry but  with redshifts
taken from alternative  surveys; and sample III, which  includes an additional
$38,672$ galaxies which do not have measured redshifts due to fiber-collision,
but were assigned the redshift  of nearest neighbors (cf.  Zehavi \etal 2002).
The present  analysis is based on  the sample II, which  consists of $369,447$
galaxies distributed over 301,237 groups  with a sky coverage of $~4514\, {\rm
  deg^2}$.  Details of  the group  finder and  the general  properties  of the
groups can be found in Yang et al.  (2007).

The halo  mass of  each group is  estimated using  the ranking of  the group's
characteristic stellar  mass, $M_{\rm stellar}$, defined as  the total stellar
masses  of  all  group  members  with  absolute  magnitude  ${^{0.1}}M_r-5\log
h\leq-19.5$.  More details of the mass estimations can be found in Yang et al.
(2007).  For those  groups  with  all members  that  have absolute  magnitudes
${^{0.1}}M_r-5\log h > -19.5$, which are not assigned halo masses in the group
catalogue, we use  the mean stellar-to-halo mass relation  obtained in Yang et
al. (2009) to  assign their halo masses. In this study,  only groups with halo
masses $M\geq  10^{11.5} \msunh$ are selected.  Note also that  in these group
catalogues the survey  edge effects have been taken into  account (Yang et al.
2007). We use only those groups  with $f_{\rm edge} \geq 0.6$, where $1-f_{\rm
edge}$ is  the fraction  of galaxies  in a group  that are  missed due  to the
survey edges.

Applying  all the  above  mentioned selection  criteria  (magnitude, mass  and
$f_{\rm  edge}$),  we have  a  total of  27,173  central  galaxies and  64,366
satellite galaxies. Here the central galaxy  is defined to be the most massive
(in  terms of  stellar  mass) galaxy  in  each group  and  other galaxies  are
satellites.  These  galaxies are  used to detect  the first kind  of alignment
signals where position angles of galaxies are not required.

However,  in  order to  study  other three  kinds  of  alignment
signals,  the position angles of the galaxies are  required. For
these studies, we keep only those  galaxies with  $\emph{b/a}<0.75$
whose  isophotal position  alngles are well defined.   Here $a$ and
$b$  are the isophotal semi-major  and minor axis lengths,
respectively.  Thus  within our final sample with  27,173 central
and 64,366 satellite  galaxies, 13,890 central and 44,219  satellite
galaxies have well measured position angles.

Finally,  note  that the  selected  galaxy  groups  contain some  interlopers,
i.e.  false  members  assigned  to  a  group. If  the  distribution  of  these
interlopers is uncorrelated (or anti-correlated) with that of the true members
of the group,  our results on the first kind of  alignment would be negatively
biased.  According  to  Yang et  al.  (2007;  2005)  the average  fraction  of
interlopers in the group is less than 20\%. We have tested the effects of such
interlopers  by   assuming  that  the  distribution  of   the  interlopers  is
uncorrelated with the shape of the  group and is spherical, and found that the
presence  of the  interlopers can  decrease the  first type  of  the alignment
signal by $\sim 10\%$.

\subsection{Galaxies: early and late types, position angles}

In our study we  follow the prescription of Park \& Choi  (2005) to divide our
sample into the early (ellipticals and lenticulars ) and the late (spirals and
irregulars) morphological types.  This division is based on  their location in
the $u  - r$ versus $g  - i $ color  gradient space, and also  in the $i$-band
concentration index  space. The  early type galaxies  are classified  as those
lying above the boundary lines  passing through the points (3.5, -0.15), (2.6,
-0.15), and  (1.0, 0.3) in  the $u-r$ versus  $\Delta(g - i)$ space.  They are
also   required   to  have   the   (inverse)   concentration  index   $c\equiv
R_{50}/R_{90}<0.43$, where $R_{50}$ and $R_{90}$ are the radii from the center
of galaxy containing $50\%$ and $90\%$  of the Petrosian flux. The rest of the
galaxies  are   classified  as   late-type  galaxies.  The   completeness  and
reliability of this  classification scheme reaches 90\%.  For  more details of
the morphology classification, we refer the reader to Park \& Choi (2005).  In
this paper,  we adopt the $r$  band isophotal position angle  for each galaxy,
which is given in the SDSS-DR4 (Adelman-McCarthy et al. 2006). We have checked
the distributions of these position angles and found them to be isotropic.

\subsection{The nearest neighbor group}

A very  important step in  our investigation is  to find the  nearest neighbor
group, so that  we can define/estimate the direction of  the tidal force. Here
we  combine the  pairwise  velocity differences  and  the projected  distances
between the central galaxies of the HG and its neighboring groups in selecting
the nearest neighbor.  Note that the {\it pairwise  velocity differences} here
refers  to  the line-of-sight  velocity  difference  between  the two  central
galaxies of  the groups.   For a  given host group,  its NNG  is found  in the
following way.

\begin{figure}[tbp]
\includegraphics[width=0.45\textwidth]{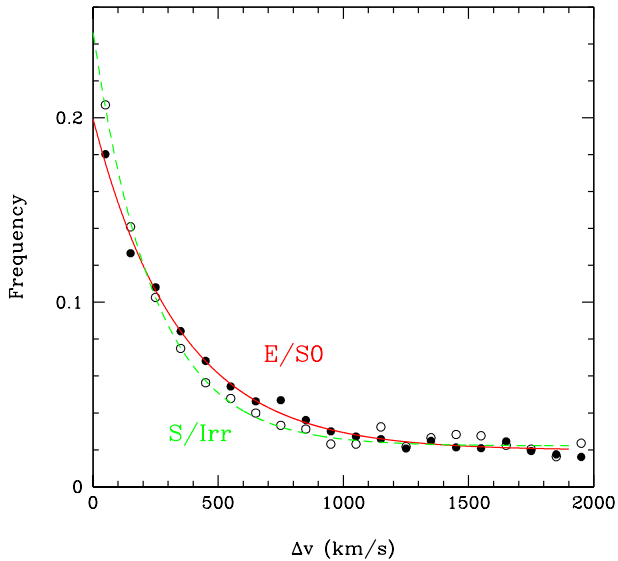}
\caption{Probability  distribution  of  the  velocity  difference  $\Delta  v$
  between early  (filled dot)  or late (open  circle) type centrals  and other
  central  galaxies at  projected separation  $r_p=0.4\sim1 \mpch$.  The solid
  line is the best fit curve for the early (E/S0) morphological type centrals,
  while the dashed one is for the late (S/Irr) type centrals.}
  \label{fig:v_d}
\end{figure}

We first inspect the distribution  of the pairwise velocity difference between
central galaxies of  the HG and the neighboring groups to  set up the velocity
difference criteria (e.g., Park  \& Choi 2009). Figure~\ref{fig:v_d} shows the
distribution of the velocity difference  $\Delta v$ for the early (filled dot)
and late (open circle) type centrals. These profiles are obtained by measuring
the velocity  difference $\Delta v$  distribution of all central  galaxy pairs
with the projected  distance $r_p$ between 0.4 and $1  \mpch$. Note that these
profiles  are contributed  by two  components: (1)  the  randomly distributed,
un-correlated pairs (constant component);  and (2) the correlated pairs (e.g.,
by neighbor  groups; enhanced  component). Since the  impact of  pairwise {\it
peculiar}  velocity  of  galaxies  is  to  broaden  the  distribution  of  the
correlated pairs, we may use the  measured profile to probe the {\it up-limit}
of the pairwise  peculiar velocity assuming that the  correlated pairs in real
space have a very compact distribution.

According   to  the  measured   $\Delta  v$   distribution  shown   in  Figure
\ref{fig:v_d},  we  fit to  the  data using  an  exponential  function plus  a
constant:
\begin{equation}
f(\Delta v) = f_1 {\rm exp}(-\Delta v/\sigma_{\Delta v})+f_2,
\end{equation}
where $f_1, f_2$ and $\sigma_{\Delta v}$ are fitting parameters.  The best fit
characteristic ({\it  up-limit}) velocity differences  $\sigma_{\Delta v}$ are
342  and 243  km/s for  early and  late type  centrals, respectively,  and the
fitting curves  are shown  as the  solid (early type)  and dashed  (late type)
lines in Figure~\ref{fig:v_d}.

We  use the following  criteria to  select the  NNG: (1)  for a
given central galaxy  of a HG,  if the  central galaxy  of the
neighbor group  has velocity difference  less than  800  (for early
type  central) or  600 (for  late-type central)  km/s (about 2.4
times the  characteristic velocity  difference) and projected
separation $r_p  < 3\mpch$, the neighbor with  the smallest $r_p$ is
set  as the  NNG;  (2)  if there  is  no central  galaxy  within the
velocity difference or  $r_p$ limit, the  neighbor with the smallest
three dimensional separation $s$ is set to as the NNG, since this
group has a higher probability of being the true NNG in real space
than the other ones with smaller $r_p$ but beyond the velocity
difference limit.  These velocity values (800 km/s and 600 km/s,
roughly 2.4 times the  characteristic velocity difference) are
chosen as a  reasonable compromise  between obtaining  a large
sample and  reducing the fraction  of  interlopers in  the  sample.
Since  the line-of-sight  pairwise peculiar velocity of galaxies  is
typically 300 km/s (see Figure~\ref{fig:v_d}), it is very unlikely
for redshift  distortion to make real space close pairs to be
separated in  redshift space with velocity difference  much larger
than 300 km/s,  so the three  dimensional distance  in redshift
space between  the two central galaxies can be used as the distance
indicator. On the other hand, the central galaxy pairs with velocity
difference less than 800 km/s (early types) or 600 km/s (late types)
and $r_p$ less than $3\mpch $  are modeled using the projected
separation $r_p$ as the distance indicator. However, we have checked
varying  these velocity  differences  (800 km/s  for early type  and
600  for late-type) and  found that our results  are not sensitive
to  the exact values used.  Finally,  note that  one can  also
define  the  NNG  according to  the projected separation relative to
the size  of the host group (e.g., its virial radius). However, we
have made  tests  and again  found that  using such  an alternative
definition of the NNG will  not have any significant impact on our
results.

\section{Quantifying the alignment}
\begin{figure*}
\begin{center}
  \includegraphics[scale=0.72,angle=270] {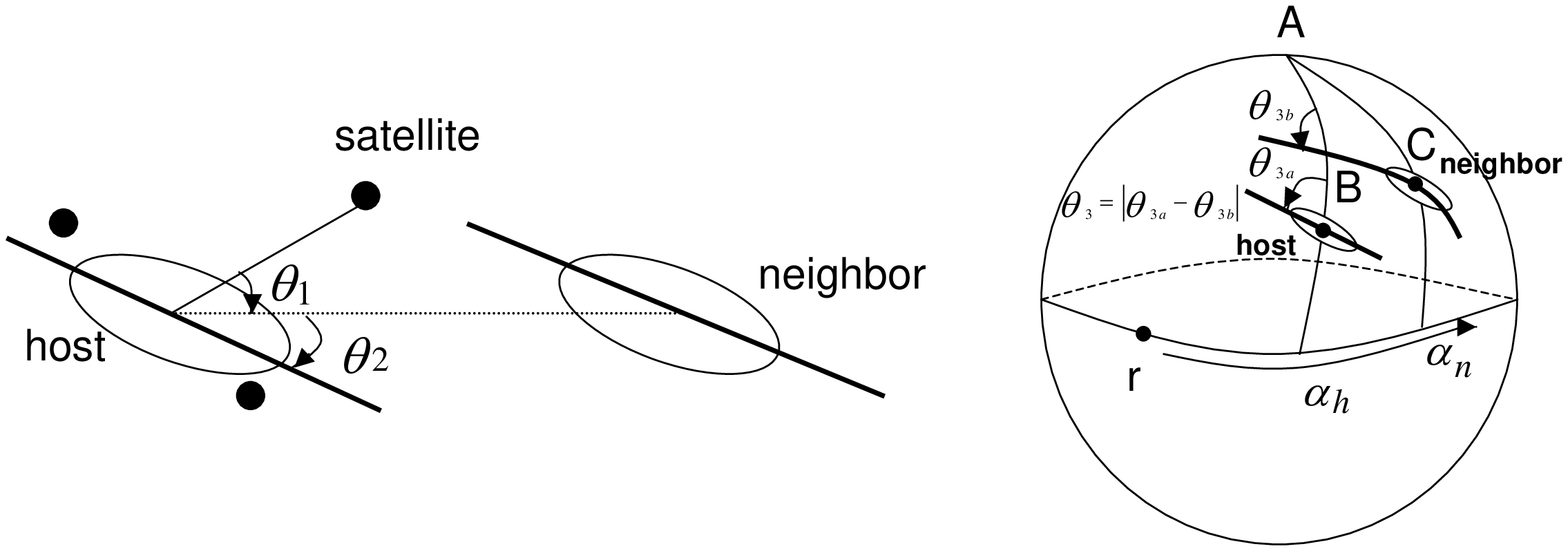}\caption{Illustration of
    the first three alignment angles $\theta_1$, $\theta_2$ (left panel) and
    $\theta_3$ (right panel).} \label{fig:ang}
\end{center}
\end{figure*}

To study the impacts  of the NNG on the various properties  of the central and
satellite galaxies,  we first  measure the different  alignment signals  as we
listed in Section 1.  The alignments  of objects are obtained by computing the
distribution  functions  of  the   alignment  angles  (e.g.,  Brainerd  2005),
$P(\theta_i)  (i=1,2,3,4)$, where  $\theta_i$  is the  angle  between the  two
directions.

The angle  $\theta_1$ is the projected  angle between the  line
connecting the central galaxy  to the  satellite galaxy and  the
line connecting  the central galaxy to  the NNG  (See the  left
panel of  Figure \ref{fig:ang}).  The angle $\theta_1$  is
constrained  in   the  range  $0^{\degr}  \leq  \theta_1  \leq
180^{\degr}$, where $\theta_1 <  90^{\degr} (>90^{\degr})$ implies a
satellite is at the near (far) side of the HG with respect to the
NNG. We also define an angle
\begin{equation}
\tilde{\theta_1} \equiv \left\{
\begin{array}{ll}
\theta_1, & \theta_1 < 90^{\degr}\\
180^{\degr} - \theta_1, & \theta_1 >90^{\degr}
\end{array}\right.
\end{equation}
The range of $\tilde{\theta_1}$ is $0^{\degr} \leq \tilde{\theta_1}
\leq 90^{\degr} $, which is more useful when making average.

The angle $\theta_2$ is the angle between the major axis of the
central galaxy of  the HG  and  the  direction of  the  NNG (See the
left  panel of  Figure \ref{fig:ang}),  which is  constrained to  be
in  the  range $0^{\degr}  \leq \theta_2 \leq90^{\degr}$. $\theta_2
= 0^{\degr}  (90^{\degr})$ suggests that the major (minor) axis of
the central  galaxy of the HG  is perfectly aligned with the
direction of the NNG.

\begin{figure}[htb]
\includegraphics[width=0.45\textwidth]{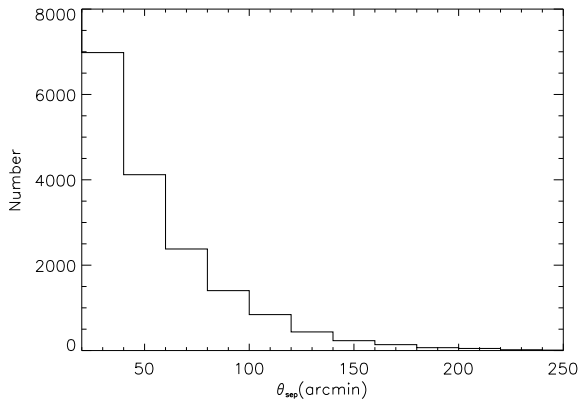}
\caption{The distribution of angular separations  between HGs and NNGs for our
  sample.  } \label{fp_sep}
\end{figure}

The angle  $\theta_3$ is the angle between  the two major axes  of
the central galaxies of the  HG and the NNG. As shown  in
Figure~\ref{fp_sep}, the angular separations  between HG  and  NNG
on  the  celestial sphere  can  reach up  to $2^{\degr}$ for some
pairs, so for  the sake of accuracy, we have included the curvature
effect in the determination  of $\theta_3$ by parallel transport the
angles  along  great  circles  on  the  celestial  sphere.  The
procedure  is illustrated in  the right panel  of
Figure~\ref{fig:ang}, and explained  in more details in  the
Appendix. Note however that  neglecting this effect  does not have
any significant  impact on  our results.  The angle  $\theta_3$  is
also constrained in the range $0^{\degr} \leq \theta_3 \leq
90^{\degr}$.

The angle $\theta_4$  is similar to $\theta_2$ but defined  for the major axes
of  the satellite  galaxies, i.e.,  the angle  between the  major axis  of the
satellite galaxy of the HG and the direction of the NNG.

In the following we will take  $\theta_1$ as an example to explain the process
of     measuring    the    alignment     signals,    the     measurement    of
$\theta_2,\theta_3,\theta_4$ are similar.

For  a  given  set  of  the  HG-NNG  pairs,  we  first  count  the  number  of
satellite-central-NNG  pairs, $N(\theta_1)$,  that have  the  angle $\theta_1$
between  the central-satellite direction  and the  direction of  the NNG  in a
number of $\theta_1$ bins.  Next, we  construct 100 random samples in which we
randomize  the  positions of  all  NNGs,  and  compute $\langle  N_R(\theta_1)
\rangle$, the  average number of satellite-central-NNG pairs  for the randomly
located NNGs as a function  of $\theta_1$. The random samples constructed this
way  suffer exactly  the same  selection effects  as the  real sample,  so any
significant  difference between $N(\theta_1)$  and $N_R(\theta_1)$  reflects a
genuine alignment between the distribution of the satellite galaxies in the HG
and the direction of the NNG.

Following  Y06  and  W08,  we  quantify  the alignment  signal  by  using  the
distribution of normalized pair counts:
\begin{equation}\label{eq:fpairs}
f_{\rm pairs}(\theta_1)=\frac{N(\theta_1)}{\langle N_R(\theta_1)\rangle}.
\end{equation}
In the  absence of any  alignment, $f_{\rm pairs}(\theta_1)=1$,  while $f_{\rm
pairs}(\theta_1) >  1$ near  $\theta =  0$ or 180  degree implies  a satellite
distribution preferentially aligned along the direction of the NNG.

We  may quantify  the fluctuation  using $\sigma_{\rm  R}(\theta_1)
/ \langle N_{\rm R}(\theta_1) \rangle$, where $\sigma_{\rm R}$ is
the standard deviation of $N_{\rm R}(\theta)$, which could  be
estimated from the 100 random samples. In addition to  this
normalized pair count, we also  compute the average angle $\langle
\theta_1    \rangle$.    In   the   absence    of   any   alignment
$\langle\tilde{\theta_1}  \rangle   =  45^{\degr}$.  If   one  finds
$\langle \tilde{\theta_1}  \rangle <45^{\degr}$  ($\langle
\tilde{\theta_1} \rangle  > 45^{\degr}$), it means that the
satellites are parallel (perpendicular) to the line connecting the
host-neighbor pairs.

\section{Results}

\subsection{Satellite galaxy distribution relative to the direction of the
  NNG}\label{sec:theta1}

As have  been found in recent  papers, the satellite  galaxies are distributed
preferentially along the major axis  of the central galaxies, especially those
with red  central galaxies (e.g.,  Brainerd 2005; Agustsson \&  Brainerd 2006;
Y06;  Azzaro  et al.  2007;  Kang  et al.  2007;  W08).  This  shows that  the
distribution  of the  satellite  galaxies  is not  completely  random, but  is
correlated  with the shapes  of the  central galaxies.  In this  subsection we
check whether similar correlations exist beyond the single dark matter halo of
the group.   In order to describe  the deviation of the  alignment signal from
the null, a $\chi^2$ test is applied here
\begin{equation}\label{eq:chi2}
\chi^2=\sum_{i=1}^{N_\mathrm{bin}}\frac{(f_\mathrm{pairs}-1.0)^2}{\sigma_\mathrm
{R}^2}
\end{equation}
where $N_\mathrm{bin}$ denotes the bin number of the angular distribution.

\begin{figure}[htb]
\includegraphics[width=0.49\textwidth]{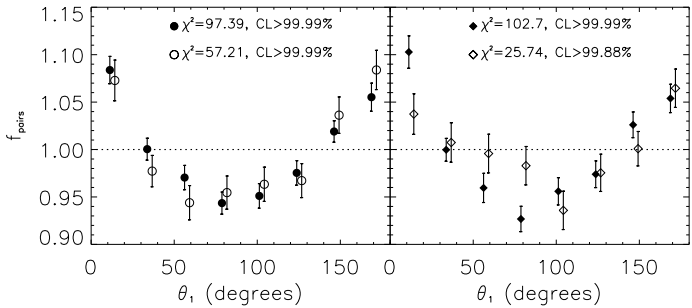}
\caption{The  normalized probability  distribution for  $\theta_1$,  the angle
  between the directions of the satellite galaxies in the HG and the NNG.  The
  left  panel shows  results  for the  whole  sample (filled  circle) and  the
  subsample with projected distance smaller than $3 r_{vir}$ of the host group
  (open circle).  The right panel shows  results for the  subsamples where the
  mass of the  HG are larger (filled diamond) and  smaller (open diamond) than
  that of the NNG.  The open symbols are shifted slightly along the horizontal
  axis for clarity.}
\label{fp_ns1}
\end{figure}

The filled  circles in  the left-hand panel  of Figure~\ref{fp_ns1}
shows the distribution of  $f(\theta_1)$ for all  satellite-NNG
pairs in our  SDSS group catalogue.   One  can see  that  $f_{\rm
pairs}(\theta_1)>1$  at small  (near $0\degr$) and large $\theta_1$
(near $180\degr$) values, while it is less than 1 at middle values
(near $90\degr$).  From the figure, we see clearly that the
distribution of HG  satellites are not completely uniform  or
isotropic, there is a  small (in absolute strength)  but highly
significant  preference for the direction along  the line connecting
HG and  NNG. The deviation  from uniform distribution has
$\chi^2=97.39$, corresponding  to  $\rm{CL}>99.99\%$ for  8 degree
of freedom  (hereafter dof) . The average  value of $
\tilde{\theta_1}$ is $\langle \tilde{\theta_1}\rangle =
43.8^{\degr}\pm0.1^{\degr}$, which again shows that the distribution
of satellites  of HG are not isotropic or uniform, but slightly
prefers the direction of the NNG.  The effect is small but highly
significant  ($\sim10\sigma$). For  comparison, the  open circle  in
the left panel  of  Figure~\ref{fp_ns1}  shows   $f_{\rm  pairs}
(\theta_1)$  for  the host-neighbor pairs  with the projected
distance  smaller than 3  times of the virial radius of the HG.
There is no significant difference between the filled circle  and
open  circle lines,  the confidence  of this  difference  is below
$0.5\sigma$ level.  {\it In other words,  not only the NNGs but also
the large
  scale environments, e.g. the filaments,  represented by the NNGs that affect
  (or at least  play an important role in) the  distribution of the satellites
  (and  the mass  distribution within  the  host halo).}   Otherwise we  would
expect distance dependent signals.

Note that in the above discussion, the  NNG can be either more
massive or less massive than  the HG,  it may  be interesting to
investigate these  two cases separately. For  this purpose, in the
right-hand  panel of Figure~\ref{fp_ns1} we show  the resulting
$\theta_1$ distributions for  HG-NNG systems  with the mass of HG
larger (filled diamond) and smaller (open diamond) than that of the
NNG. In  the rest part  of the paper,  unless stated otherwise, we
use filled circle symbol to represent the sample which is not
constrained by the distance limit  while the  open circles  to
represent  the results  with  the projected distance smaller than 3
times the  virial radius of the HG; the filled diamond (open
diamond) symbol represents the  results for subsamples where the
mass of the HG are larger (smaller) than  that of the NNG. The
stronger alignment signals shown  in the right-hand panel in
Figure~\ref{fp_ns1} for satellite galaxies in more massive HGs may
caused by the fact that their distributions are flatter  than those
in smaller HGs  (e.g., Jing  \& Suto 2002;  Yang et al. 2006).

\begin{figure}[tb]
\includegraphics[width=0.5\textwidth]{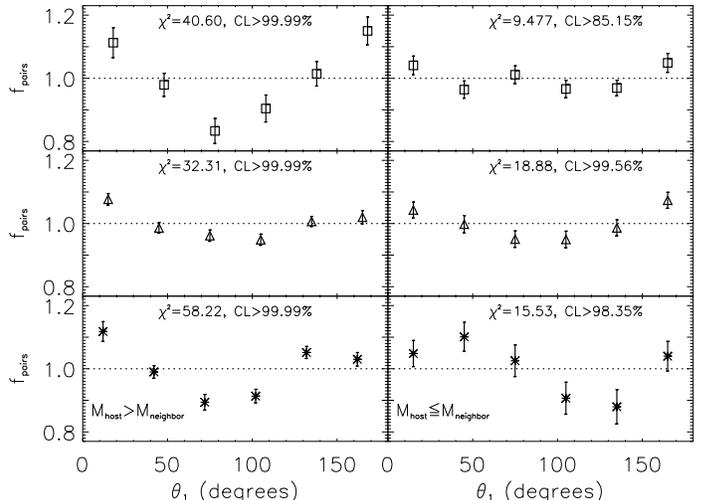}
\caption{Similar   to  Fig.~\ref{fp_ns1},  but   here  for   different  HG-NNG
  systems. Left panels: results for HG-NNG systems where the mass of the HG is
  larger than that of the NNG.  Right panels: results for HG-NNG systems where
  the mass of the HG is smaller  than that of the NNG.  The asterisk, triangle
  and  square symbols represent  the mass  of the  NNG in  the range  $M_n \in
  [11.5,12.5]$,           $[12.5,13.5]$           and           $[13.5,14.5]$,
  respectively. } \label{fp_mass_alig1}
\end{figure}

We  can further  check  this by  measuring  the alignment  signals
for  HG-NNG systems that  are divided  into subsamples of  different
masses.  In  the left panel of  Figure~\ref{fp_mass_alig1}, we show
the alignment  signal of $f_{\rm pairs}(\theta_1)$ for the submaples
with mass of the HG  that are larger than that of the NNG, while in
the right  panel of the Figure~\ref{fp_mass_alig1}, we show  the
results of the  subsamples with the  mass of the HG smaller than
that of  the NNG.  In  each panel, the  asterisk, triangle and
square symbols represent the mass  of the NNG in the range $M_n
\equiv \log_{10} (h M/\Msun) \in [11.5,12.5]$, $[12.5,13.5]$ and
$[13.5,14.5]$, respectively. We found some interesting    trends
here.    According    to    the    left panel    of
Figure~\ref{fp_mass_alig1},  for  the most  massive groups  ($M_n
>13.5$),  the alignment  signal (i.e.   deviation of $f_{\rm
pairs}(\theta_1)$  from  1) is strongest ($5\sigma$), perhaps
indicating the  very strong attraction  by the NNG  and the flatter
distribution of  satellite galaxies  within the  HG. For system with
less massive HG and  NNGs ($M_n \in [12.5,13.5]$),  the signal is
weaker(deviation from  null signal hypothesis by $3\sigma$),
However, for the smallest groups  discussed here ($M_n \in
[11.5,12.5]$),  the alignment signal is again rather  strong
(deviation from null signal  hypothesis by 5$\sigma$). This  is also
evident from  the  value  of  $\chi^2$, 58.22,  32.31,  40.60,
corresponding  to the  mass of  the NNG  in the  range $M_n  \in
[13.5,14.5]$, [12.5,13.5] and  [11.5,12.5], respectively.
Interestingly,  for those systems with NNG  heavier than HG  and NNG
in  the range $M_n\in  [11.5,12.5]$ (bottom right panel), there are
slightly  more satellite galaxies distributed near the NNG (with
$\theta_1 < 90\degr$) than HG (the confidence level is $1.5\sigma$),
which may  indicate that  the NNG near  a small  HG may attract  its
satellite galaxies and affect their distribution.

\begin{figure}[tb]
\includegraphics[width=0.5\textwidth]{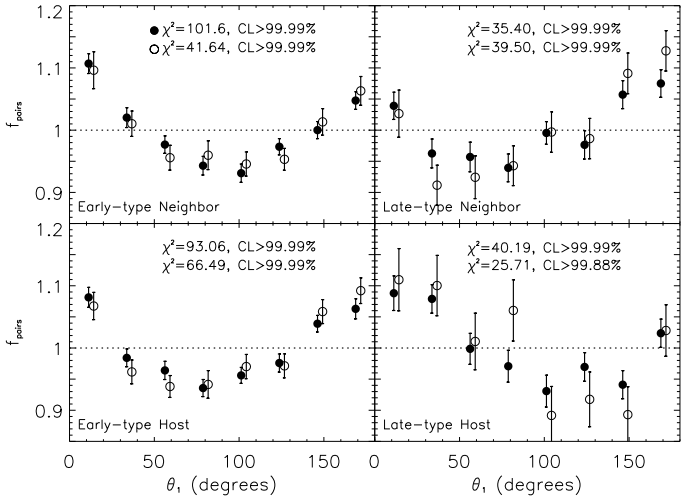}
\caption{Same as
    Fig.~\ref{fp_ns1}, but for different subsamples, divided by the type of
    the central galaxies of the HG or NNG.} \label{fp_ns2}
\end{figure}

\begin{figure}[tb]
\includegraphics[width=0.5\textwidth]{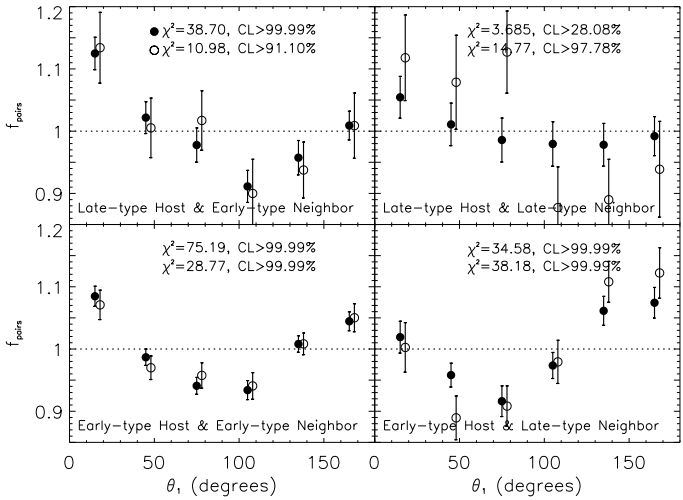}
\caption{Same as
    Fig.~\ref{fp_ns2}, except that here we split the sample according to the
    morphological types of both the central galaxies of the HG and
    NNG. } \label{fp_ns3}
\end{figure}

In order to study how this alignment depends on the morphological types of the
central galaxies of the  HG and the NNG, following Y06 and  W08, we divide our
sample  of host  and  neighbor groups  into  different morphology  subsamples.
Figure~\ref{fp_ns2} shows the  alignment signals $f_{\rm pairs}(\theta_1)$ for
HG  with early (lower-left  panel) and  late-type (lower-right  panel) central
galaxy,  and NNG  with  early (upper-left  panel)  and late-type  (upper-right
panel) central galaxy, respectively.

The alignment  signal does  seem to  depend on the  morphological type  of the
central galaxy slightly:  the groups with early type  central galaxies or with
the NNGs having an early  central galaxy show slightly stronger alignment than
those with late-type  centrals ($1.0\sigma$ for the early  centrals of HG than
late  centrals of  HG, $1.7\sigma$  for the  early centrals  of NNG  than late
centrals  of NNG).  In  Figure~\ref{fp_ns3}, we  show  the alignment,  $f_{\rm
pairs}(\theta_1)$,  for four  combinations  of  the HG  and  NNG with  central
galaxies of  different morphological  types.  As one  can see, pairs  with the
late-type HG  and late-type  NNG show the  smallest strength of  the alignment
signal.

The HG satellite distribution is  also slightly asymmetric with
respect to the near  side or  far  side  from the  NNG.   There is a
small but  significant preference for  the satellite  galaxies in a
group with  a late  type central galaxy to be distributed  near the
side of the NNG, with  either an early or a late type central
galaxy.  This asymmetry is significant by $\sim 2\sigma$ for both
the whole  sample and  and for  the case  where the  NNG  with
projected separation is smaller than 3 times  of the virial radius.
For the samples with the  early-type HG  and late-type  NNG,
however,  the trend  is  opposite: the satellite galaxies  are
preferentially  distributed on the  far side  from the NNG, again
this effect is  small but significant ($2.4\sigma$ for all samples,
$2.7\sigma$ for the close pairs). The  reason of this is not
completely clear, but it  may be due  to (i) the  groups with late
type central galaxy  is more probably located in the outskirts of  a
high density region with groups having preferentially early type
central  galaxies; (ii) probably smaller than groups with early type
central galaxies, thus have smaller impact on its NNG.  Before we
proceed, it is quite interesting to check whether or not the weak
asymmetry is induced by those groups near the survey edges. As we
have tested using only groups with  $f_{\rm edge}>0.9$ by performing
exactly the same  analysis, the weak asymmetry is  almost the same.
Moreover, one would  only expect that the groups near the edge may
slightly induce the weak  asymmetry in the direction of the {\it
near} side of the NNG,  however not in the {\it far}  side of the
NNG, as shown in the lower-right  panel of Figure~\ref{fp_ns3}.
Therefore, we believe that our results (e.g., the weak asymmetry)
are robust against the impact of groups near the survey edges.

\begin{figure}[tbp]
\includegraphics[width=0.5\textwidth]{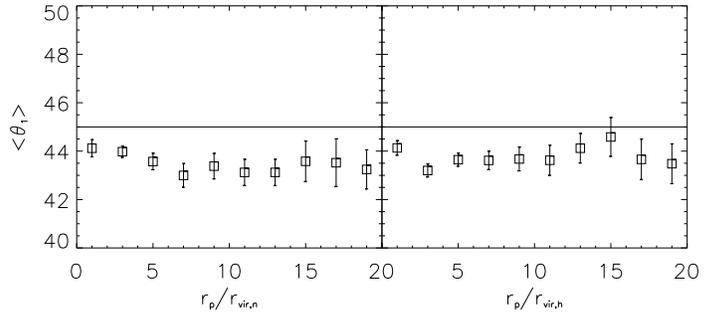}
\caption{The average value of $\theta_1$ as a function of the projected
  distance between the HG and the NNG.  Left: as a function of $r_{vir, n}$;
  Right: as a function of $r_{vir, h}$.} \label{fp_r1}
\end{figure}

To what  extent does  the large  scale environment as  represented by  the NNG
affect the  distribution of the satellite  galaxies in the HG?   To study this
problem,  we  divided  the  total  sample into  subsamples  according  to  the
projected distance between  the HG and the NNG.   Figure~\ref{fp_r1} shows the
dependence of the  average value $\theta_1$ on the  projected distance between
the HG and the  NNG.  Here we take both the virial  radius of the NNG, $r_{\rm
  vir, n}$(left  panel), and the  HG, $r_{\rm vir,  h}$ (right panel),  as the
unit. There is  no significant difference between the results  in the left and
right panels.  The  value of $\langle \theta_1 \rangle$  depends weakly on the
distance between  the HG  and NNG. The  alignment between the  distribution of
satellite galaxies and  the direction of the NNG  is significant (greater than
$1\sigma$) up to  separations as large as about  12$r_{\rm vir, n}$(or $r_{\rm
  vir,h}$), which is also evident from the fact that $\langle \tilde{\theta_1}
\rangle  = 43.1\degr  \pm 0.5\degr$  (or $43.6\degr  \pm 0.6\degr$ )
for the systems with  separation of the  HG and the  NNG being
12$r_{\rm vir,  n}$ (or $r_{\rm vir, h}$). From these alignment
signals for $\theta_1$ at so large separations, we conclude that the
distributions of the satellite galaxies in HGs are (also) affected
by the large scale environments.

\subsection{The position angle of the central galaxy}

\subsubsection{relative to the direction of the NNG}

Both Y06  and W08 found that  there is a  preference for the satellites  to be
distributed along the major axis of  their central galaxy, and in the previous
section, we found a prominent  alignment signal between the orientation of the
satellite system of the HG and the direction of the NNG.  Therefore, we expect
there is also an alignment between the major axis of the central galaxy of the
HG and the direction of the NNG.

\begin{figure}[tbp]
\includegraphics[width=0.5\textwidth]{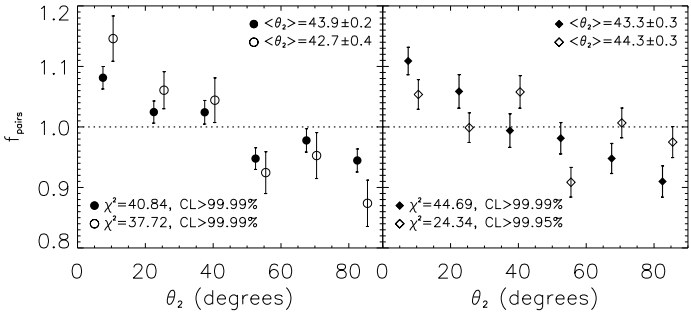}
\caption{The  normalized probability distribution,  $f_{\rm pairs}(\theta_2)$,
  of the angle $\theta_2$ between the  major axes of the central galaxy of the
  HG and the direction of the NNG. In the left panel, results are measured for
  all samples (filled  circle) and the subsamples with  the projected distance
  smaller than 3  times the virial radius of the host  group (open circle). In
  the right panel, we show results for subsamples where the mass of the HG are
  larger (smaller)  than that  of the NNG  with filled diamond  (open diamond)
  symbol. Formal rejection confidence levels  from the $\chi^2$ test are shown
  in each panel. The open circles and open diamonds have been shifted slightly
  along the horizontal axis for clarity.} \label{fp_chn1}
\end{figure}

Indeed, as shown in the left  panel of Figure~\ref{fp_chn1}, there
is a strong signal  of alignment  between  these two  directions:
the major  axis of  the central galaxy of  the HG is preferentially
aligned with  the direction of the NNG.   The  alignment signal  is
even stronger  for  the  pairs with  smaller projected   separations
as   shown   by   the   open   circle   symbols   of
Figure~\ref{fp_chn1}.  This result suggests that the major axis of
the central galaxy  of the  HG tends  to be  aligned  with the
direction of  the NNG  (or possibly the large scale structure
beyond).

The orientation of the central galaxy of HG is possibly affected by
(1)the potential of the HG; (2)NNG and (3) the large scale
environment. To understand the relative influence of these factors,
we checked separately the signals for all pairs of NNG and the close
pairs ($\le 3r_{vir}$), and find the following average values:
$\langle \theta_2 \rangle = 43.9\degr \pm 0.2\degr$ ($\langle
\theta_2 \rangle = 42.7\degr \pm  0.4\degr$ for the close pairs),
which depend quite significantly on  the pair separations. Thus it
is unlikely for the large scale environment to be main factor in
determining the orientation of the central galaxy, otherwise one
would expect $\theta_2$ to be insensitive to the distance. On the
other hand, in W08 it was found that the orientation of the central
galaxy is also aligned with the potential of the HG (provided that
the satellite  distributions trace  the  mass distribution of the
halo reasonably well). Therefore, we conclude both the
 the satellite distribution (or the mass
distribution) in the  HG and  the NNG at small  separation can
affect the orientation  of the central galaxy. However, from the
alignment signal measured for $\theta_1$, we find that the
distribution of the satellite is affected by both the NNG and the
large scale of environment (see the last part of section 4.1), so
there may be indirect correlation between the orientation of the
central galaxy and the large scale environment.

In  the right-hand   panel  of Figure~\ref{fp_chn1}  we show the
resulting $\theta_2$ distributions  for HG-NNG systems with the mass
of HG larger (filled diamond) and smaller (open diamond) than that
of the NNG.   The position angle of the central galaxy in more
massive host group is  slightly more aligned (by $1.4\sigma$) which
is very likely caused by its flatter distribution of the satellite
galaxies.

\begin{figure}[tbp]
\includegraphics[width=0.5\textwidth]{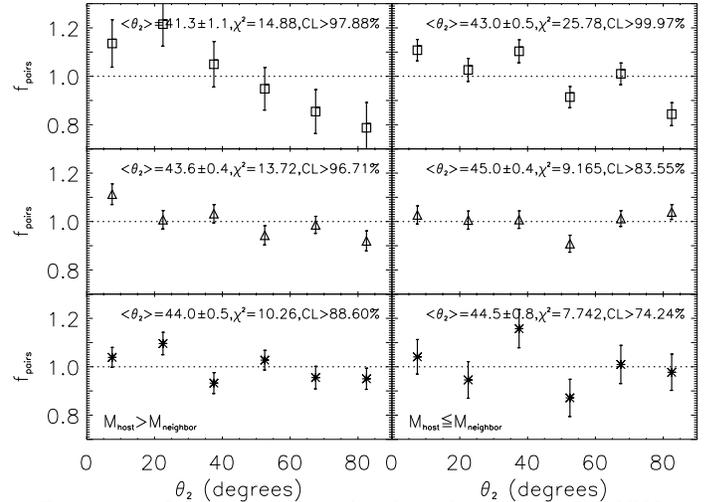}
\caption{Similar  to   Fig.~\ref{fp_chn1},  but  here   for  different  HG-NNG
  systems. Left panels: results for HG-NNG systems where the mass of the HG is
  larger than that of the NNG.  Right panels: results for HG-NNG systems where
  the mass  of the  HG is smaller  than that  of the NNG.  In each  panel, the
  asterisk, triangle  and square points represent  the mass of the  NNG in the
  range    $M_n   \in    [11.5,12.5]$,   $[12.5,13.5]$    and   $[13.5,14.5]$,
  respectively. }
\label{fp_mass_alig2}
\end{figure}


To examine the  dependence of the alignment signal  on the group
mass, we divide our  sample into three subsamples according to the
NNG mass as in Section \ref{sec:theta1}.  Figure~\ref{fp_mass_alig2}
shows the distribution of  $\theta_2$   for  systems   with  the NNG
mass  in  the   range  $M_n \in[13.5,14.5]$, $[12.5,13.5]$, and
$[11.5,12.5]$ from the top to the bottom panel.  The panels on the
left are  for the systems having a HG more massive than  its  NNG,
and  those  on  the  right  are  for  the  opposite  cases.
Figure~\ref{fp_mass_alig2}  clearly  shows  that  the  alignment
signal  is stronger  for more  massive systems.   The statistical
significance  of the alignment is $3.3\sigma,  3.5\sigma$, and
$2.0\sigma$ from top  to bottom in the left column.  It is
$4.0\sigma, 0.0\sigma$, and $0.6\sigma$ in the same order in the
right column. The top panels tell that the groups in a pair are
aligned  with  each  other  when  the  NNG  mass  exceeds
$10^{13.5}\msunh$ regardless of  the HG  mass. Such  alignment can
occur  by the  strong tidal force of the  NNG on the central galaxy
of the HG. This signal  due to NNGs disappears when the  NNG mass is
less than $10^{13.5}\msunh$  as can be seen in the middle  right and
bottom right panels.  Instead, the alignment signal is seen when the
HG mass is higher than the NNG  mass, which means that the NNG is
now aligned,  regardless of its  mass, along  the major axis  of the
central galaxy of the HG.

\begin{figure}[tb]
\includegraphics[width=0.5\textwidth]{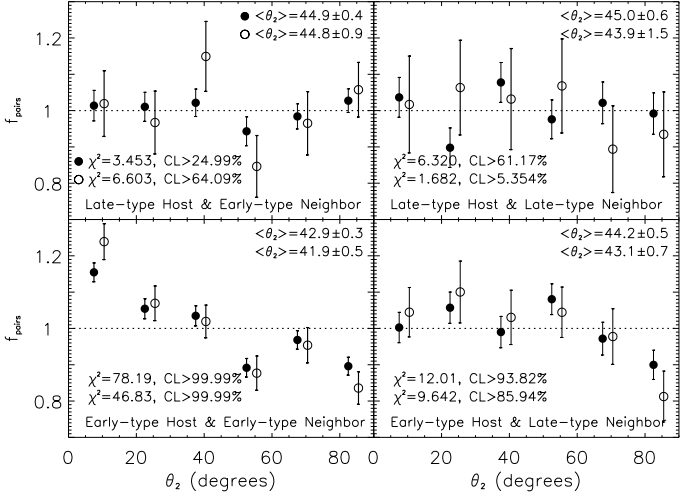}
\caption{Same as Figure~\ref{fp_chn1}, but for different subsamples of central
  galaxies of the host and neighbor groups. } \label{fp_chn2}
\end{figure}

In  Figure~\ref{fp_chn2},  we  examine how  $f_{\rm
pairs}(\theta_2)$ depends on  the morphological  type of the central
galaxies. Note  that the early  types are  mainly ellipticals,  and
the  elongation of  an early-type galaxy can be due to  external
gravitational effects or internal kinematics, both of which are
closely  correlated with the distribution of the satellite galaxies
and the NNGs. On the other hand, the late types are dominantly disk
galaxies  and the  position angle  of  a disk  galaxy is  determined
by  the direction of  its spin axis. Therefore, for HG central
galaxies with different morphological types, the alignment with
respect to the NNG could have different physical origins. In the
bottom two panels of  Fig.~\ref{fp_chn2}, we show the  alignment
signals $f(\theta_2)$ for HGs  with early-type centrals. The left
panel is for the HGs  whose NNG has an early-type central while the
right is for those whose NNG contains a late-type  central.  The HGs
having an early-type central show  quite a  strong alignment signal,
particularly when the central galaxy of the NNG is also an early
type (a $7\sigma$ effect for  a sample of  all such  pairs and
$6\sigma$ for  close pairs). If the central galaxy of the NNG is a
late type one, the alignment signal weakens considerably, the
deviation from the case of no alignment is only $1.6\sigma$ and
$2.7\sigma$ for all pairs and close pairs, respectively (see the
lower-right panel of Figure~\ref{fp_chn2}).


\begin{figure}[tbp]
\includegraphics[width=0.5\textwidth]{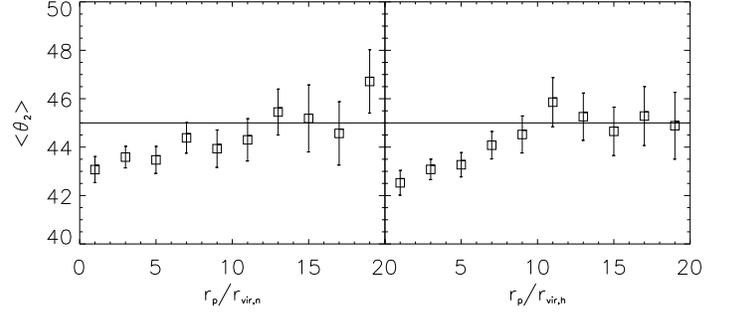}
\caption{The average value of $\theta_2$ as a function of the projected
  distance between the HG and the NNG. Left: as a function of $r_{vir, n}$;
  Right: as a function of $r_{vir, h}$.  } \label{fp_r2}
\end{figure}

In the case where the central galaxy of  the NNG is a late type one,
the major axis of  the galaxy is not  so well correlated with  its
stellar distribution, but more with the  inclination of the disk. On
the other  hand, the minor axis is quite well correlated with the
disk spin axis. Thus $\theta_2=0$ means that the  spin  axis   is
perpendicular  to  the  direction   of  the  NNG,  while
$\theta_2=90^{\degr}$ means  that the spin axis  tends to be aligned
with the direction of the NNG. In the upper panels of
Figure~\ref{fp_chn2}, we show the distribution $f(\theta_2)$ for the
late type central galaxies (Upper Left: NNG with early type central;
Upper Right: NNG with late type central). However, we do not find
significant alignment signal between the major axis of the central
late type galaxy and the direction of  the NNGs. As we have also
measured the  average alignment  angle $\langle  \theta_2 \rangle$
as a  function of distance  between the  HG and  NNG  as shown  in
Fig.~\ref{fp_r2},  $\langle \theta_2  \rangle$ increases  slowly and
approaches  the  null position  of $45\degr$ as the distance
increases.  The results shown in Fig.~\ref{fp_r2} indicate  that
only  NNGs at  separation  $\la 5r_{vir}$  may have  possible
impacts onto the orientations of the central galaxies in the HGs.

\subsubsection{relative to the orientation of the central galaxy of the NNG}

\begin{figure}[tb]
\includegraphics[width=0.5\textwidth]{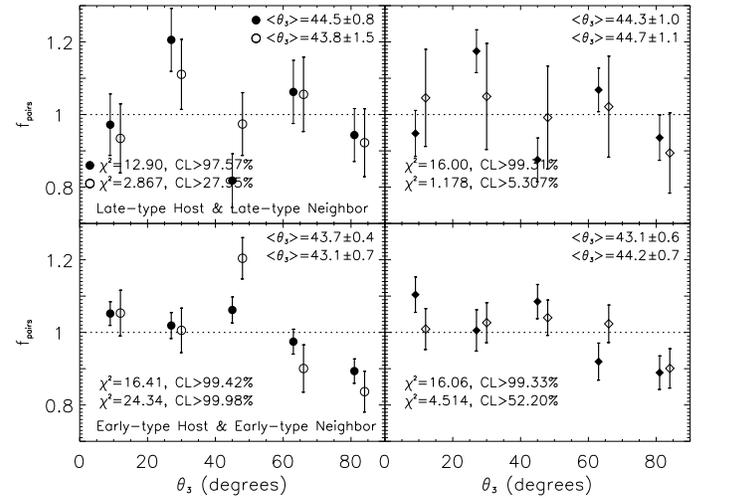}
\caption{The normalized probability distribution, $f_{\rm pairs}(\theta_3)$,
  of the angle $\theta_3$ between the two major axes of the central galaxies
  of the HG and the NNG. Upper panels: the central galaxies of the HG and NNG
  are both late types. Lower panels: the central galaxies of the HG and NNG
  are both early types. In the two left panels we show results for the whole
  sample (filled circle) and the subsample with the projected distance smaller
  than 3 times the virial radius of the host group (open circle). In the two
  right panels, we show the results for subsamples where the mass of the HG
  are large (smaller) than that of the NNG with filled diamond (open diamond)
  symbol.} \label{fp_nh1}
\end{figure}

In Figure~\ref{fp_nh1},  we present the  distribution of the  angle
$\theta_3$ between the major axes of the central  galaxy of the HG
and the central galaxy of the  NNG. In  the lower panels  we use
only the HG  and NNG  whose central galaxies are both early types.
One can  see that the major axes of the central galaxies of the HG
and the NNG tend to be parallel, though the signal is weak. We also
checked  the case of alignment between all  types of central
galaxies, and find there is no prominent  signal, as shown by the
fact $\langle \theta_3 \rangle  =  44.6\degr \pm  0.3\degr$
(deviates from  the  null  case by  only $1.3\sigma $).  However,
the $\theta_3$ alignment signal for  the case of the HG more massive
than the NNG is stronger than the opposite case by $1.8\sigma$ (see
the  lower-right panel). Combined  with our probe  of the alignment
signals  for  $\theta_1$  and  $\theta_2$,  we find  that  the large
scale environments tend to impact the  distribution of satellite
galaxies (and the mass distribution) in the HGs,  while the
distribution of satellite galaxies and possibly  NNGs at  small
separations may  impact the orientations  of the (early-type)
central galaxies  in the HGs. Thus the  alignment signals shown in
$\theta_3$ for early type galaxies are expected.

The upper panels of  Figure~\ref{fp_nh1} shows the distributions of $\theta_3$
when the  morphological types of  the central galaxies  of the HG and  NNG are
both  late  types.  The  spin  axis  of the  two  galaxies  are  parallel  for
$\theta_3=0\degr$  and perpendicular  if  $\theta_3 =90\degr$.  We find  large
fluctuations in the distribution, and there is no significant alignment signal
between the  spin axes of  the late-type central galaxies,  regardless whether
the HG is more massive or less than the NNG.

\subsection{The position angle of the satellite galaxy relative to the
  direction of the NNG}

Many  studies  have attempted  to  detect  the  alignment signal
between  the orientations of  central galaxies  and satellite
galaxies  in the  clusters of galaxies  (e.g., Plionis et  al. 2003;
Strazzullo et  al.  2005;  Torlina, De Propris \& West 2007), most
of them found only null or weak signal. Using a sample of 4289
host-satellites pairs from the SDSS  DR4, Agustsson \& Brainerd
(2006) found a weak signal of  the alignment. Adopting the same
group  catalogue as that used here,  Faltenbacher et  al.  (2007)
searched  for the alignment between  the orientations of the BGG and
the satellites. They considered the total sample and submaples with
different color of  the satellite galaxy, and found a small but
definite  alignment signal between the  major axes of the  central
and the satellite galaxies  of   the  host   group,  especially at
small  scales $r_p<0.1r_{\rm vir,h}$, where  one expects a strong
tidal  force at such small separation.

\begin{figure}[tb]
\includegraphics[width=0.5\textwidth]{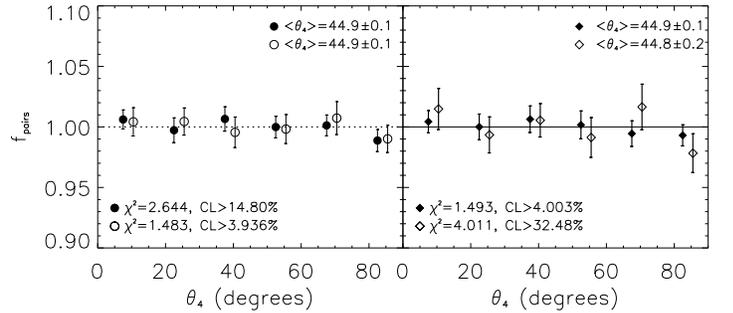}
\caption{The  normalized probability distribution,  $f_{\rm pairs}(\theta_4)$,
  of the angle $\theta_4$ between  the orientations of the satellite galaxy in
  HG and the  direction linking the host and NNG.  In  the left panel, results
  are measured  for the whole sample  (solid line) and the  subsample with the
  distance  limit,  3  times the  virial  radius  of  the host  group  (dotted
  line). In the right panel, we show results for the subsamples where the mass
  of the HG are large (smaller)  than that of the NNG with dashed (dot-dashed)
  line.} \label{fp_sat1}
\end{figure}

\begin{figure}[htb]
\includegraphics[width=0.5\textwidth]{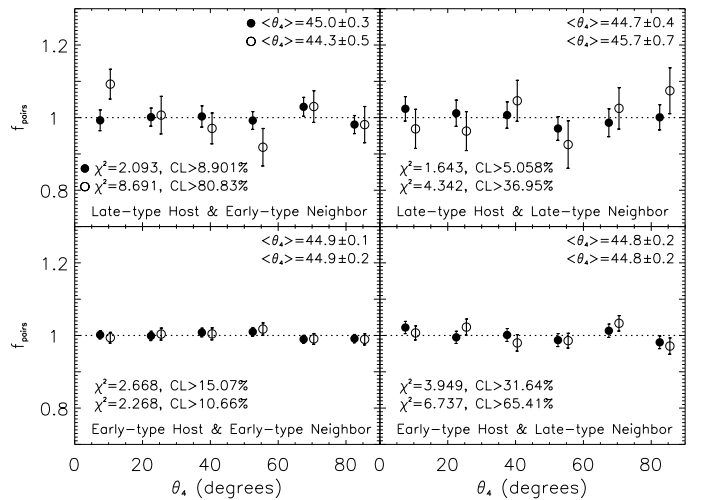}
\caption{Same  as  Figure~\ref{fp_sat1},   but  for  subsamples  of  different
  morphological  types   of  central  galaxies   of  the  host   and  neighbor
  groups. } \label{fp_sat2}
\end{figure}

As we have already noticed, the NNG and the large scale environment can affect
the distribution of  satellite galaxies in the HG.  Here, we  check if the NNG
(and  the large  scale environment)  can also  impact the  orientation  of the
satellite galaxies.   The method to  obtain the alignment angle  $\theta_4$ is
similar to that for the angle  $\theta_3$, the only difference is that here we
use  the major axes  of the  satellites to  substitute the  major axis  of the
central galaxy of the HG.   Fig.~\ref{fp_sat1} shows the alignment between the
major axes of the satellites in the HGs and the direction linking the host and
NNGs. There is apparently no alignment between the major axes of the satellite
and the  direction linking the  host and the NNG.   Fig~\ref{fp_sat2} displays
the dependence on  the $f_{\rm pairs}(\theta_4)$ on the  morphological type of
the central galaxies  of the HGs and NNGs. Again we  do not find statistically
significant alignment signal for the morphology subsamples.  Moreover, we also
checked the  distribution $f_{\rm  pairs}(\theta_4)$ separately for  the early
and late type satellite galaxies, but there is no significant alignment signal
either.

\section{Summary and discussion}

In the cold dark matter scenario,  small dark matter halos form first and grow
subsequently  to larger  structures via  accretion and  merger  processes. The
accretion  of  material  might  preferentially  occur  along  the  filamentary
structure  (West  1994), which  leads  to  correlations  between the  internal
structures of  the neighboring groups.  Binggeli (1982)  pioneered the studies
of the alignment between neighboring  clusters of galaxies, and found that the
host clusters tend to be  aligned with their nearest neighbors. The subsequent
studies  confirmed  this   tendency,  although  conflicting  results  appeared
occasionally in the  literature (West 1989; Plionis 1994).   In this paper, we
used  the host-neighboring  group systems  extracted from  the SDSS  DR4 group
catalogue  (Yang  et  al.  2007)  to  probe  the  impact  of the  large  scale
environment  (as  represented  by   the  nearest  neighboring  group)  on  the
distribution of the  satellite galaxies and on the  orientation of the central
and satellite galaxies.  For this purpose, four types of alignment signals are
measured and the main results are summarized as follows.

\begin{enumerate}
\item  There is  a strong  alignment signal  between the  distribution  of the
  satellites  relative to  the direction  of the  NNG. This  signal  is rather
  insensitive  to  the separation  between  the HG  and  NNG,  and extends  to
  separation beyond  $12r_{vir}$ of  the HG.  For  the system with  both early
  central galaxies of  the HG and NNG, the alignment  signal is the strongest,
  for the  system with both  late central galaxies,  the signal is  weaker but
  still quite significant.

\item  The  satellite  galaxies  in  the  HG have  a  weak  preference  to  be
  distributed at the  near side of the NNG with an  early type central galaxy,
  and at the far side of the NNG with a late type central galaxy.

\item The  major axis  of the  central galaxy of  the HG  is aligned  with the
  direction  of  the NNG,  especially  in the  massive  HGs.   This effect  is
  stronger for  the systems when the central  galaxies of HG and  NNG are both
  early types.  And  we find this alignment signal only  exists between HG and
  NNG pairs at separation $\la 5r_{vir}$ of the HG.

\item The  distribution of the satellite  galaxies and the  orientation of the
  central galaxy  of the HG  show stronger alignment  signals with the  NNG in
  systems with more massive HGs and NNGs.

\item There is a preference for the  two major axes of the central galaxies of
  the HG  and NNG to be  parallel for the  system with the both  early central
  galaxies, while there  is no evident correlation between  the two major axes
  of the central galaxies for the systems with both late type cental galaxies.

\item Although the distribution of the satellites of the HG is correlated with
  the  direction of  the NNG,  their  orientations (position  angles) are  not
  correlated with the direction of the NNG.
\end{enumerate}

According to our various alignment measurements, we find that the distribution
of  satellite galaxies and  the orientation  of the  central galaxy  both show
strong alignment signals with respect  to the direction of the NNG ($\theta_1$
and $\theta_2$).  Such  alignment signals are stronger in  massive halos where
central  galaxies  are  early  type  ones.   Because of  these  two  kinds  of
alignments, the alignment  between the orientations of the  cetral galaxies of
the HG and  the NNG ($\theta_3$) is naturally  expected.  For the orientations
of satellite  galaxies, however, we  do not find significant  alignment signal
relative  to the  direction of  NNG  ($\theta_4$), while  Faltenbacher et  al.
(2007) have found prominent alignment  signals between the orientations of the
central and  satellite galaxies at  very small separation.  This  may indicate
that the orientations of satellite  galaxies are only strongly affected by the
tidal force of the central galaxy  and the host halo, but not significantly by
the NNG or the large scale environment.

Possible explanations for the strong alignment signals in the first and second
types of the alignments  are: (i) the NNG affects the shape  of the host halo,
while  the distribution of  satellite galaxies  and the  shape of  the central
galaxy are mainly affected by the  host halo; (ii) the large scale environment
directly affects the  distribution of satellite galaxies and  the shape of the
central galaxy.  Judging from the first alignment signal, it seems likely that
the impact of large scale environment is  mainly on the shape of the host halo
instead of  directly on  the distribution of  the satellite  galaxies, because
otherwise we should expect to find  a strong dependence on the distance to the
NNG, rather than a mass dependence of  the HG and NNG.  The fact that stronger
alignment signals is  found for the subsamples with more  massive HGs and NNGs
is likely  correlated with the  fact that the  more massive halos  have larger
triaxialities (e.g, Jing \& Suto 2002; Wang \& Fan 2004; W08).

\begin{figure}[tbp]
\includegraphics[width=0.5\textwidth]{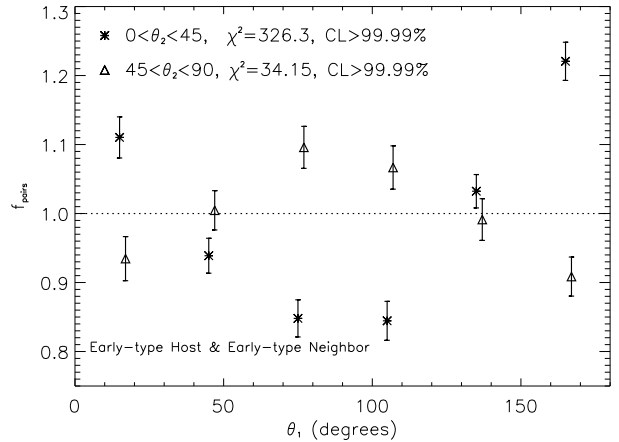}
\caption{The  normalized probability distribution,  $f_{\rm pairs}(\theta_1)$,
  of the angle $\theta_1$ for  the systems with early-type host and early-type
  neighbor.  The asterisk  and triangle symbols represent the  results for the
  submaples  with $0^{\degr}<\theta_2  < 45^{\degr}$  and $45^{\degr}<\theta_2
  <90^{\degr}$, respectively.  The triangle symbols have been shifted slightly
  along the horizontal axis for clarity.  } \label{fp_alig1}
\end{figure}

In  order  to  check  whether  the  shape of  the  central  galaxy
is  mainly determined/affected by its  own host halo, or by the  NNG
at small separation, we  carried out  an  additional  test.  In
Fig.~\ref{fp_alig1},  we show  the alignment signal $f_{\rm
pairs}(\theta_1)$ for the systems with the early-type host and
early-type neighbor, where  the asterisk and triangle symbols show
the results for  the system with  $0^{\degr}<\theta_2 < 45^{\degr}$
and $45^{\degr}<\theta_2  <90^{\degr}$, respectively. When
$45^{\degr}<\theta_2 <90^{\degr}$,  the major axis  of the central
galaxy of  the  HG is  rather perpendicular to the direction of the
NNG. If the shape  of central galaxy is not much affected by the
host  halo, but only  affected by the NNG,  we would expect that the
alignment  signals of  the two  subsamples are similar. The results,
however,  show that  the alignment signals are opposite for the two
different subsamples (the difference level is $7.5\sigma$). On the
other hand, the signals are precisely what one would expect if the
mass and light are reasonably  well-aligned (i.e., the  image of
early type  central galaxy and the surrounding mass  of the HG) and
the satellites in  the HG trace the surrounding mass. That is, one
would get a  peak at $\theta_1 = 90^{\degr}$ if $45^{\degr} <
\theta_2 < 90^{\degr}$ and a valley at $\theta_1 = 90^{\degr}$ if
$0^{\degr} < \theta_2 < 45^{\degr}$.   Hence, the shape and
orientation of  the central galaxy  is more likely  to  be
determined  by  its  own host  halo.   While  the  different
separation   dependencies   of   $\langle   \theta_2   \rangle$
shown   in Figure~\ref{fp_r2}  and $\langle \theta_1  \rangle$ shown
in  Figure~\ref{fp_r1} indicate  that  NNG at  small  separation may
also  play,  however, only  a secondary  impact on  the orientation
of the  central galaxy  of the  HG at particular conditions, e.g,
interaction with the host halo (e.g., Lin et al. 2003; Ludlow et al.
2009). This conclusion is also strengthened by the recent
measurements which  show that the orientations of  the central
galaxies are preferentially  aligned with the distribution of
satellite galaxies (e.g. Y06 and references therein).

Finally, we draw our conclusion that the large scale environment traced by the
NNG have  impacts on the shape (orientation)  of the host halo.   On the other
hand, the  distribution of the satellite  galaxies, the shapes  of the central
galaxies, and the shape of the satellite galaxies at small radii, however, are
mainly affected by their own host  halos.  Apart from these, the NNG also have
direct impacts  on the  distribution of satellite  galaxies, which  produce an
asymmetric alignment signal  with respective to the near or  far away sides of
the NNG.  And NNG at small separations may also have small secondary impact on
the orientation of the central galaxy of the HG.


\section*{Acknowledgments}

We  thank  the  referee  for   the  constructive  and  detailed
comments  and suggestions. This  work has started during  YGW's
visit to KIAS,  and he would like to express  his gratitude for
KIAS.  CBP and  YYC acknowledge the support of  the   Korea  Science
and  Engineering  Foundation   (KOSEF)  through  the Astrophysical
Research  Center for the  Structure and Evolution of  the Cosmos
(ARCSEC).  XY  acknowledges the support  by the Shanghai Pujiang
Program (No. 07pj14102),  973 Program  (No.   2007CB815402), the CAS
Knowledge  Innovation Program  (Grant  No.  KJCX2-YW-T05)  and
grants  from  NSFC (Nos.   10533030, 10673023, 10821302). YGW  and
XLC acknowledges the support  of the 973 program (No.2007CB815401),
the   CAS  Knowledge   Innovation   Program  (Grant   No.
KJCX3-SYW-N2), and the NSFC grant 10503010. YGW is also supported by
the Young Researcher Grant of National Astronomical Observatories,
Chinese Academy of Sciences. XLC is also supported by the NSFC
Distinguished Young Scholar Grant No.10525314.

Funding for  the SDSS  and SDSS-II has  been provided  by the Alfred  P. Sloan
Foundation, the  Participating Institutions, the  National Science Foundation,
the   U.S.  Department  of   Energy,  the   National  Aeronautics   and  Space
Administration, the  Japanese Monbukagakusho, the Max Planck  Society, and the
Higher  Education  Funding  Council  for   England.   The  SDSS  Web  Site  is
http://www.sdss.org/.

The  SDSS  is  managed  by  the  Astrophysical  Research  Consortium  for  the
Participating  Institutions. The Participating  Institutions are  the American
Museum  of Natural  History,  Astrophysical Institute  Potsdam, University  of
Basel, Cambridge  University, Case  Western Reserve University,  University of
Chicago, Drexel  University, Fermilab, the  Institute for Advanced  Study, the
Japan Participation  Group, Johns Hopkins University, the  Joint Institute for
Nuclear  Astrophysics,  the  Kavli  Institute for  Particle  Astrophysics  and
Cosmology,  the  Korean  Scientist  Group,  the Chinese  Academy  of  Sciences
(LAMOST),  Los  Alamos   National  Laboratory,  the  Max-Planck-Institute  for
Astronomy (MPIA), the Max-Planck-Institute  for Astrophysics (MPA), New Mexico
State University, Ohio State  University, University of Pittsburgh, University
of Portsmouth, Princeton University,  the United States Naval Observatory, and
the University of Washington.

\appendix

In this appendix, we explain how  to measure the angle between the
orientation of the central galaxy of the host group and NN, taking
into account the effect of  curvature  of  the  sky.   We  take  one
pair  of  the  host  $(\alpha_h, \delta_h,\Phi_h)$  and neighbor
$(\alpha_n,  \delta_n,\Phi_n)$ as  an example, where  $\alpha_{h}$,
$\delta_h$  and  $\Phi_h$  are   the  right  ascension, declination,
and  the position  angle of the  host centrals  respectively, and
$\alpha_{n}$, $\delta_n$  and $\delta_n$  are the corresponding
parameters of the NN. Here we assume that $\alpha_n>\alpha_h$
$\delta_h>0$, and $\delta_n>0$ (See the  right panel of
Figure~\ref{fig:ang}  ). Extending the  major axis of the central
galaxy of  the NN, it will cross the longitude  of the host at the
point  B. Now  in the  spherical triangle  $\bigtriangleup ABC$,  we
have two angles $\angle BAC$  and $\angle ACB$ and one  side
$\widehat{AC}$. Therefore, we  can  get  the  angle  $\angle  CBA$
by  solving  the  spherical  triangle $\bigtriangleup ABC$.  It  is
clear that the angle  between the orientation of the  central galaxy
of the  host group  and NN  can be  written as  $\Phi_h -
(180^{\degr}  - \angle CBA)$,  which is  equivalent to  the angle
$\theta_3 = \theta_{3a}  - \theta_{3b}$  (see  the right  panel of
Figure~\ref{fig:ang}). Position angles of the major axes are with
respect to the east direction.

\end{document}